\documentclass[12pt]{iopart}
\usepackage{amssymb}
\usepackage[usenames]{color} 

\usepackage{graphicx}
\usepackage{bm}
\newcommand{\abs}[1]{\ensuremath{\left\vert#1\right\vert}}
\newcommand{\ket}[1]{\ensuremath{\vert#1\rangle}}
\newcommand{\bra}[1]{\ensuremath{\langle #1\vert}}

\newcommand{\kb}[2]{\ensuremath{\vert #1 \rangle \langle #2 \vert}}
\newcommand{\exx}[1]{\ensuremath{\langle #1 \rangle}}

\renewcommand{\sp}[0]{\ensuremath{\mathbf{\sigma}_{+}}}
\newcommand{\sm}[0]{\ensuremath{\mathbf{\sigma}_{-}}}
\newcommand{\id}[0]{\ensuremath{\mathbf{1}}}

\newcommand{\an}[1]{\ensuremath{#1}}
\newcommand{\cre}[1]{\ensuremath{#1^\dagger}}

\newcommand{\ddt}[0]{\frac{\mathrm{d}}{\mathrm{d}t}}
\renewcommand{\vec}[1]{\ensuremath{\mathbf{#1}}}

\begin{document}

\title{Robust adiabatic approach to optical spin entangling in coupled quantum dots}
\author{Erik M Gauger}
\ead{erik.gauger@materials.ox.ac.uk}
\address{Department of Materials, University of Oxford, OX1 3PH, United Kingdom}
\author{Ahsan Nazir}
\address{Department of Physics and Astronomy, University College London, Gower Street, London WC1E 6BT, United Kingdom}
\address{Centre for
Quantum Dynamics and Centre for Quantum Computer Technology, Griffith
University, Brisbane, Queensland 4111, Australia}
\author{Simon C Benjamin}
\address{Department of Materials, University of Oxford, OX1 3PH, United Kingdom}
\address{Centre for Quantum Technologies, National University of Singapore, 3 Science Drive 2, Singapore 117543}
\author{Thomas M Stace}
\address{Department of Physics, University of Queensland, Brisbane 4072, Australia}
\author{Brendon  W Lovett}
\ead{brendon.lovett@materials.ox.ac.uk}
\address{Department of Materials, University of Oxford, OX1 3PH, United Kingdom}
\date{\today}

\begin{abstract}
Excitonic transitions offer a possible route to ultrafast optical spin manipulation in coupled nanostructures. We perform here a detailed study of the three principal exciton-mediated decoherence channels for optically-controlled electron spin qubits in coupled quantum dots: radiative decay of the excitonic state, exciton-phonon interactions, and Landau-Zener transitions between laser-dressed states. We consider a scheme to produce an entangling controlled-phase gate on a pair of coupled spins which, in its simplest dynamic form, renders the system subject to fast decoherence rates associated with exciton creation during the gating operation. In contrast, we show that an adiabatic approach employing off-resonant laser excitation allows us to suppress all sources of decoherence simultaneously, significantly increasing the fidelity of operations at only a relatively small gating time cost. We find that controlled-phase gates accurate to one part in $10^2$ can realistically be achieved with the adiabatic approach, whereas the conventional dynamic approach does not appear to support a fidelity suitable for scalable quantum computation. Our predictions could be demonstrated experimentally in the near future.
\end{abstract}

\maketitle

\section{Introduction}

Quantum computation (QC) promises a significant speedup for certain classes of problems \cite{nielsen00, shor97, grover96} as well as an efficient way of simulating quantum systems \cite{abrams97}. The broad range of expertise and knowledge that exists in fabricating and characterising semiconductor structures has naturally led to a number of proposals for quantum dot (QD) QC being put forward. The spin of a single electron confined within a QD has been suggested as a scalable qubit~\cite{loss98}, while another possibility is to define the computational basis states as the presence or the absence of a confined electron-hole pair (an exciton). While excitonic qubits can be conveniently optically addressed, they also interact strongly with their surroundings and thus suffer from rapid decoherence (typically within a nanosecond~\cite{borri01}).
Spin-based qubits, on the other hand, have much longer coherence times (up to a millisecond \cite{kroutvar04}) but can suffer from the fact that inter-spin interactions tend to be weak. 
Hybrid schemes have therefore emerged that propose to store the qubit in a long-lived localized electronic spin state, moving selectively to an excited state spin-exciton (trion) representation only when a gate is actually performed. Such schemes hope to marry the advantages of the two different representations~\cite{calarco03, imamoglu99}, though it should be borne in mind that any excitonic component of the system state will still suffer from rapid decoherence, and hence significant excited state population is ideally avoided. 

For universal QC it suffices to have arbitrary single qubit operations and at least one entangling two qubit gate \cite{nielsen00}. Single qubit operations are simply rotations of the state vector on the Bloch sphere and proposals based both on stimulated Raman adiabatic passage and direct optical excitation exist for spins in QDs~\cite{bergmann98, roszak05, chen04}. Decoherence processes have been thoroughly studied in these cases~\cite{roszak05, grodecka07, caillet07, gauger08}, leading to the prediction that high-fidelity single-qubit operations should be feasible. In many physical systems, the easiest entangling operation to implement is a controlled-phase (CPHASE) gate that leaves three of the four possible two-qubit computational basis states unaltered, while generating a phase of $\pi$ on the fourth (e.g. $\ket{00}\rightarrow\ket{00}$, $\ket{01}\rightarrow\ket{01}$, $\ket{10}\rightarrow\ket{10}$, $\ket{11}\rightarrow-\ket{11}$) \cite{nielsen00}. Various schemes to realise such a gate exist for spins in coupled QDs~\cite{calarco03, nazir04, biolatti02}, of which adiabatic optical control is a particularly promising example as it naturally suppresses excited state population throughout the operation \cite{calarco03, lovett05}. The technique relies on varying external control parameters of the system, here the intensity and frequency of a laser, sufficiently slowly compared to the characteristic timescales of the system itself that the entire population remains in instantaneous eigenstates throughout the gating operation. This approach overcomes light-heavy-hole-mixing in the QD valence band \cite{calarco03, lovett05} and is also thought to be robust to phonon-induced dephasing mechanisms~\cite{calarco03}.
While the adiabatic approach seems extremely promising as perhaps the best way to achieve rapid control of solid state spins, it does have potential issues like every approach. It is therefore essential to perform a comprehensive study of all major noise sources during a {\em two-qubit} gate. Here we provide such a study, and conclude that the approach is in fact remarkably robust. 
We address the principal decoherence channels of radiative recombination and carrier-phonon interactions by deriving a Markovian master equation (ME) for the laser-driven two-dot system, accounting for both processes. We will use the ME to assess the performance of the adiabatic approach within experimentally accessible parameter regimes. We also derive conditions to avoid non-adiabatic Landau-Zener (LZ) transitions between eigenstates \cite{landau32, zener32, stueckelberg32, wubs05}, an effect that poses an additional error source. In order to put the results in context, we will consider a simple dynamic approach based on resonant Rabi flopping as a comparison.

\section{The System}\label{model}

By observing Rabi oscillations, optical coherent control of excitons in single QDs has been demonstrated by a number of groups in recent years \cite{stievater01, zrenner02, ramsay07}. Further, excitonic interactions between QDs have been observed experimentally~\cite{kagan96, berglund02, crooker02} and optical conditional logic between two excitonic qubits within a single QD has been accomplished~\cite{li03}. 
QD spin qubit Rabi flopping has been achieved both through direct spin resonance (on the sub-microsecond time scale)~\cite{koppens06} and optically (on the picosecond time scale)~\cite{berezovsky08}. Single-shot read-out by spin to charge conversion \cite{elzerman04} has also been achieved. In addition, Greilich et al.~\cite{greilich06b} have performed optical measurements of spin coherence and Xu et al~\cite{xu07} have done fast spin cooling using excitons. 

We consider a realistic state-of-the-art system: Two adjacent self-assembled QDs (I and II respectively) with a distinct heavy- and light-hole valence band structure, such as those grown on a GaAs substrate~\cite{bayer99}. The dots could be placed side-by-side, though are more usually grown in stacks (see Ref. \cite{xie95} for example). We consider QDs that are small and have a strong confinement potential that dominates over any intra-dot Coulomb interactions. They are each doped such that a single excess electron, the spin of which embodies the qubit, permanently occupies the lowest energy state of the conduction band. The qubit basis is defined as $\ket{0}\equiv\ket{\downarrow}\equiv\ket{-1/2}$ and $\ket{1}\equiv\ket{\uparrow}\equiv\ket{1/2}$, where $\ket{m_z}$ is the spin projection, with $m_z=\pm1/2$. The $z$ direction also defines the quantization axis for the hole spins (see below). Our qubit representation is different to that used by Imamoglu {\it et al.}~\cite{imamoglu99}, where an applied magnetic field defines the
qubit basis in the $x$-direction. Further, since our scheme relies on
direct inter-dot interactions rather than the cavity mediated Raman
transitions needed in Ref.~\cite{imamoglu99}, we require only a single
off-resonant laser pulse to effect a CPHASE gate.

\begin{figure}
\begin{center}
\includegraphics[width=0.75\textwidth]{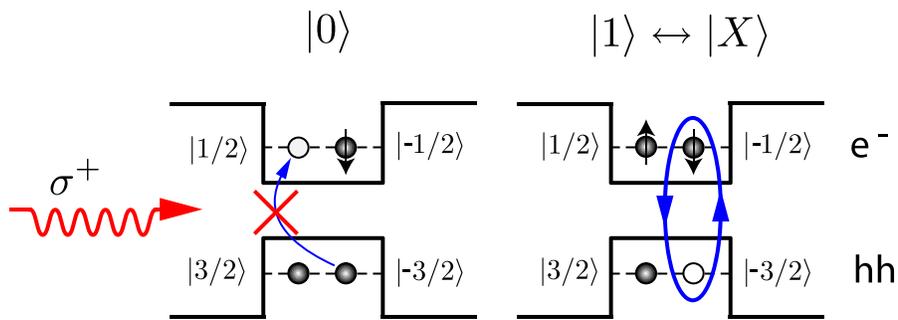}
\caption{Pauli blocking effect: $\sigma^+$ light cannot excite electron hole pairs if the qubit is in the state $\ket{0}$ with $m_z = -1/2$ (left hand side). On the other hand, if the qubit is in the $\ket{1}$ configuration the excitation is possible, leading to a three particle trion state $\ket{X}$ (right hand side).}
\label{fig:pauli_blocking}
\end{center}
\end{figure}

In order to couple the two spin qubits we exploit the fact that trion-trion inter-dot Coulomb interactions are expected to be relatively strong \cite{pazy03,nazir04} such that we can address the system optically through spin-selective exciton creation~\cite{calarco03,pazy03, nazir04}. This effect, which has recently been observed experimentally \cite{chen00, yokoi05}, relies on the Pauli exclusion principle for the lowest energy conduction band electron state (see Fig.~\ref{fig:pauli_blocking}). This state can be occupied by at most two electrons, each of opposite spin orientation, with the electron qubit permanently occupying one of the slots. In the valence band, the heavy and light hole energy levels split due to their differing effective masses. Heavy holes occupy the lowest energy states with spin $m_z = \pm 3/2$ (whereas the light holes have spin $m_z = \pm 1/2$) and are the only valence band states considered here (see Ref.~\cite{lovett05} for detailed calculations of the effects of light-heavy hole mixing). A $\sigma^+$ circularly polarized laser incident on the QD  carries photons with angular momentum $l = + 1$. Thus, if the qubit is in state $\ket{0}$, with spin $m_z = -1/2$, promoting another electron to the conduction band is incompatible with the conservation of angular momentum and no transition can occur. Conversely, for a qubit in state $\ket{1}$ ($m_z=1/2$) the transition is allowed and leads to the creation of a three-particle trion state $\ket{X}\equiv\ket{\uparrow, \downarrow, \triangledown}$, where the arrows symbolize the spin projection of the electron, and the triangle of the hole ($\ket{\triangledown}\equiv\ket{-3/2}$), respectively.

That Coulomb interactions between trions on the two adjacent dots mediate a coupling of the two spin qubits can be seen by considering the full system Hamiltonian.
We assume that the qubit states are degenerate and write the Hamiltonian of the two QDs and a single classical laser field in the basis $\{ \ket{0}, \ket{1}, \ket{X} \}$ ($\hbar = 1$) as
\begin{eqnarray}
H(t) &=& \omega_0 (\kb{X}{X} \otimes \id + \id \otimes \kb{X}{X})\nonumber\\
&&\:{+}\Omega \cos \omega_l t(\kb{1}{X} \otimes \id + \id \otimes \kb{1}{X} + \mathrm{h.c.})  \nonumber \\
&&\:{+} V_{XX} \kb{XX}{XX} + V_F (\kb{1X}{X1} + \mathrm{h.c.} ).
\label{eqn:9ls_hamiltonian}
\end{eqnarray}
where $\mathrm{h.c.}$ denotes the Hermitian conjugate, $\bm{1}$ is the identity operator, and $\omega_0$ is the exciton creation energy (assumed the same for both dots). Here, $\Omega$ represents the coupling between the QD transition and laser mode and $\omega_l$ is the laser frequency, both of which may be time-dependent quantities. The dots are Coulomb coupled both by virtual photon exchange (F\"orster interaction) of strength $V_F$ ~\cite{foerster59}, and by a biexcitonic dipolar coupling $V_{XX}$~\cite{pazy03}. We consider typical inter-dot separations of 5 -- 10 nm, close 
enough for both $V_F$ and $V_{XX}$ to be on milli-electronvolt scale. We also assume that we are in a regime where tunneling processes are suppressed~\cite{nazir05}.

To lowest nonzero order, the F\"orster coupling element is equivalent to the interaction of two point dipoles situated on dots I and II \cite{lovett03},
\begin{equation}
V_F = \frac{C}{\epsilon_r R^3} \left( \langle \vec{r}_\mathrm{I} \rangle \cdot  \langle \vec{r}_\mathrm{II} \rangle - \frac{3}{R^2} \left( \langle \vec{r}_\mathrm{I} \rangle \cdot \vec{R} \right) \left( \langle \vec{r}_\mathrm{II} \rangle \cdot \vec{R} \right)  \right),
\end{equation}
where $\vec{R}$ is the vector connecting the centre of the two dots, $\langle \vec{r}_\mathrm{I/II} \rangle$ is the position operator between the electron and hole on either dot I/II respectively,  $\epsilon_r$ is the dielectric constant, and $C$ the usual Coulomb term $C = e^2 / 4 \pi \epsilon_0$. The sign of $V_F$ is determined by the relative orientation of $\langle \vec{r}_\mathrm{I} \rangle$ and $\langle \vec{r}_\mathrm{II} \rangle$ with respect to $\vec{R}$. Ultimately, the orientation of  $\langle \vec{r}_\mathrm{I/II} \rangle$ depends on the lattice vector structure \cite{lovett03} and is hard to predict. However, for two identical dots in the same crystal structure, $\langle \vec{r}_\mathrm{I} \rangle$ and $\langle \vec{r}_\mathrm{II} \rangle$ should be parallel and have the same magnitude, such that their joint orientation to $\vec{R}$ is the sole factor determining whether $V_F$ is positive or negative. While the sign of $V_F$ is an immutable property of any given two-dot system, both positive and negative values of $V_F$ should be possible in principle by a suitable geometrical arrangement.

It can be seen from Eq. (\ref{eqn:9ls_hamiltonian}) that the Hamiltonian decouples into four non-interacting subspaces~\cite{nazir04}: $\mathcal{H}_0 = \{ \ket{00} \}$, $\mathcal{H}_1 = \{ \ket{01}, \ket{0X} \}$, $\mathcal{H}_{1'} = \{ \ket{10}, \ket{X0} \}$ and $\mathcal{H}_2 = \{\ket{11}, \ket{1X}, \ket{X1}, \ket{XX} \}$; this is a direct result of the Pauli-blocking effect. To implement the CPHASE operation, we need to achieve a net phase shift of $\pi$ on the input state \ket{11}, and are therefore primarily interested in the dynamics of $\mathcal{H}_2$. When we do need to consider dynamics in the two level subspaces $\mathcal{H}_1$ and $\mathcal{H}_{1'}$ we can appeal to the results of Ref.~\cite{gauger08}, which analyzed an isomorphic Hamiltonian structure.

Proceeding as in \cite{nazir04, kolli06}, we first transform $\mathcal{H}_2$  into the basis of its eigenstates in the absence of driving ($\Omega = 0$): \ket{11},  $\ket{\psi_+} = (\ket{1X} + \ket{X1})/\sqrt{2}$, $\ket{\psi_-} = (\ket{1X} - \ket{X1})/\sqrt{2}$ and \ket{XX}, giving
\begin{eqnarray}
H_2(t)	&=& (\omega_0 - V_F) \kb{\psi_-}{\psi_-} + (\omega_0 + V_F) \kb{\psi_+}{\psi_+}\nonumber\\
&&\:{+}\sqrt{2} \Omega \cos \omega_l t (\kb{11}{\psi_+}  + \kb{\psi_+}{XX} + \mathrm{h.c.}) \nonumber \\
&&\:{+} (2 \omega_0 + V_{XX}) \kb{XX}{XX}.
\label{eqn:h2sub_ham_diag}
\end{eqnarray}
We now move to a frame rotating with the laser frequency $\omega_l$, detuned from the $\ket{11} \leftrightarrow \ket{\psi_+}$ transition by an amount $\Delta = \omega_0 + V_F - \omega_l$. Within the rotating wave approximation (RWA), Eq. (\ref{eqn:h2sub_ham_diag}) becomes
\begin{eqnarray}
H_2^{'}	&=& (\Delta - 2 V_F) \kb{\psi_-}{\psi_-} + \Delta  \kb{\psi_+}{\psi_+}
		+(2 \Delta -2 V_F + V_{XX}) \kb{XX}{XX}  \nonumber \\
		&&\:{+} \frac{\Omega}{\sqrt{2}}  (\kb{11}{\psi_+} +  \kb{\psi_+}{XX} + \mathrm{h.c.}).
\label{eqn:h2sub_ham_rwa}
\end{eqnarray}
Ref.~\cite{nazir04} describes how to perform a dynamic CPHASE operation by applying a resonant $2 \pi$ laser pulse to the $\ket{11} \leftrightarrow \ket{\psi_+}$ transition (i.e. with $\Delta=0$), 
generating the required $\pi$ phase shift on state $\ket{11}$. In this case, certain conditions on the driving strength $\Omega$ must be satisified in order to suppress unwanted transitions to $\ket{XX}$ and transitions within $\mathcal{H}_{1}$ and $\mathcal{H}_{1'}$~\cite{nazir04}. However, the gate may also be operated adiabatically without those same constraints by slowly switching an off-resonant laser beam, forcing the system to follow its instantaneous eigenstates~\cite{lovett05}. The energy difference between states $\ket{11}$ and $\ket{\psi_+}$ then allows for a phase accumulation on the $\ket{11}$ state, as can be seen by analysing the eigenstates in more detail.

\begin{figure}
\begin{center}
\includegraphics[width=0.6\textwidth]{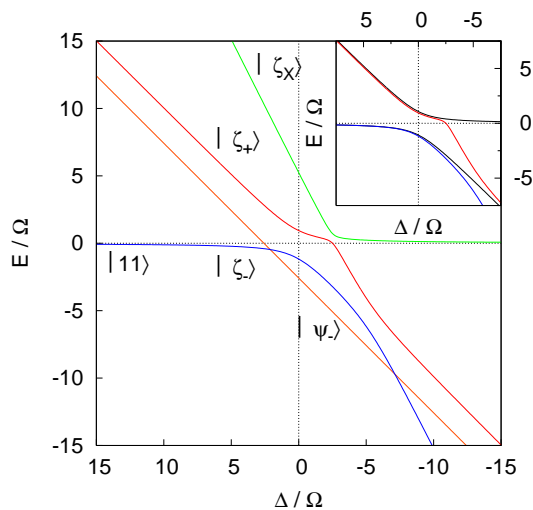}
\caption{Dressed states of subsystem $\mathcal{H}_2$.  Parameters are: $V_{XX} = 5$~meV, $V_F  = 0.85$~meV, and $\Omega = 1$~meV. The inset shows the states \ket{\zeta_-} and \ket{\zeta_+} around the origin. Also shown are the dressed states obtained if \ket{XX} is neglected. For positive $\Delta$, these approximate \ket{\zeta_-} and \ket{\zeta_+} well.}
\label{fig:dressed_states_energy_h2}
\end{center}
\end{figure}

The three levels \ket{11}, \ket{\psi_+} and \ket{XX} are coupled by the laser in Eq. (\ref{eqn:h2sub_ham_rwa}).  Moving to the diagonal basis results in three dressed states, which we shall label $\{ \ket{\zeta_-}, \ket{\zeta_+}, \ket{\zeta_X} \}$, each of which are superpositions of the three bare basis states weighted according to certain mixing angles that are complicated functions of the system parameters.
Nevertheless we can greatly simplify matters by an appropriate choice of parameters, and work with an approximate 2LS that describes the full dynamics with a high degree of accuracy. As shown in the inset of Fig. \ref{fig:dressed_states_energy_h2}, for positive $V_F = 0.85$~meV, $V_{XX} = 5$~meV, and positive $\Delta$, the energies of states \ket{\zeta_-} and \ket{\zeta_+} are very well approximated by taking 
\begin{eqnarray}\label{eqn:zetaapproxdressed}
\ket{\zeta_-} &{}\approx{}& \cos \Theta \ket{11} - \sin \Theta \ket{\psi_+}\nonumber\\
\ket{\zeta_+} &{}\approx{}& \sin \Theta \ket{11} + \cos \Theta \ket{\psi_+},
\end{eqnarray}
where $\Theta = (1/2)\arctan (\sqrt{2} \Omega / \Delta)$. The agreement improves even further for larger $\Delta$, which we shall demonstrate is desirable for the adiabatic scheme, or smaller driving $\Omega_0$. A negative F\"orster coupling $V_F < 0$ simply shifts the dark state \ket{\psi_-} upwards in Fig. \ref{fig:dressed_states_energy_h2}. At the same time, the avoided crossing between \ket{\zeta_+} and \ket{\zeta_X} moves further to the right and hence interactions with the biexcitonic level are suppressed even more effectively. Therefore, we can safely assume that the perturbation caused by \ket{XX} is negligible for a wide range of parameters.

Let us now consider shining a chirped laser on the QD with temporal evolution $\Delta(t)$ and $\Omega(t)$ that slowly and continuously changes $\Theta$ from $0$ to some value $\Theta_{max}$, and back to $0$ again. Since $\ket{11}$ and $\ket{\zeta_-}$ coincide for $\Theta = 0$, any population in state $\ket{11}$ will adiabatically follow the instantaneous eigenstate of $\ket{\zeta_-}$ and return to $\ket{11}$ at the end of the pulse. As a consequence of its time evolution, the phase of an eigenstate $\ket{\mu}$ changes according to $\ket{\mu (t)} = \exp (- i E_\mu t) \ket{\mu(0)}$, where $E_\mu$ is the relevant eigenenergy. Hence, at the end of the pulse, $\ket{11}$ picks up a phase of $\exp (-i \phi_{11})$ relative to \ket{00}, where
\begin{equation}
\phi_{11} = \int \limits_0^T \mathrm{d}t E_{\zeta_-}(t)  =  \int \limits_0^T \mathrm{d}t \frac{1}{2} \left[ \Delta(t) - \sqrt{\Delta^2(t) + 2\Omega^2(t)} \right],
\end{equation}
to a good approximation.
Here, $T$ is the pulse duration and $E_{\zeta_-}(t)$ the energy of \ket{\zeta_-} relative to \ket{00}. Of course, states $\ket{01}$ and $\ket{10}$ also accumulate phase relative to $\ket{00}$ due to the laser-induced coupling to $\ket{0X}$ and $\ket{X0}$, giving
\begin{equation}
\phi_{01,10}  =  \int \limits_0^T \mathrm{d}t \frac{1}{2} \left[ \Delta'(t) - \sqrt{\Delta'^2(t) + \Omega^2(t)} \right],
\end{equation}
respectively, where $\Delta'(t)=\Delta(t)-V_F$ is the detuning in the $\mathcal{H}_{1}$ subspace.
To achieve a gate locally equivalent to the CPHASE, $\phi_{00}-\phi_{01}-\phi_{10}+\phi_{11}=\pi$ following the operation. This constrains the control parameters $\Delta(t)$, $\Omega(t)$, and $T$.

For adiabatic following to hold, we must consider two further points. First, $\Theta$ must change sufficiently slowly that LZ transitions between the dressed states do not occur; we will consider these in Section~\ref{section:LZ}.
Second, we must start and end with $\Theta = 0$. In principal, both the laser detuning $\Delta(t)$ and the coupling strength $\Omega(t)$ can be time-dependent, though it is essential to start with an off-resonant pulse and to slowly switch the coupling strength. It is then possible, although by no means necessary, to chirp the pulse and tune into resonance~\cite{calarco03}. In this case the approach to resonance dominates the dynamics and the gating time of the operation. For the sake of simplicity and practicality, we will only consider here laser pulses with a constant detuning and Gaussian pulse envelopes. The duration of the pulse is then described by the parameter $\tau$:
\begin{eqnarray}
\Delta(t) &\equiv& \Delta  \label{eqn:pulse_detuning} \\
\Omega(t) &=& \Omega_0 e^{- (t/\tau)^2}, \label{eqn:pulse_coupling}
\end{eqnarray}
in contrast to previous work \cite{calarco03, lovett05} where chirped laser pulses were studied. Having described, in some detail, the internal dynamics of the driven system we now turn to the impact of the external environment on the CPHASE operation.

\section{Radiative decay}

The lifetime of excitons in a QD can be as long as a nanosecond \cite{kroutvar04}, while coherent control should be possible on the picosecond timescale~\cite{unold05}. However, in order to avoid phonon induced pure dephasing~\cite{alicki04} and to maintain adiabaticity it can be advantageous to perform operations much more slowly. In this case the finite exciton lifetime becomes an important source of decoherence which cannot be neglected.

Any population in an excited state of the system is susceptible to radiative decay at the rate of the inverse natural lifetime $\Gamma$, due to coupling with the vacuum radiation field. 
As discussed above, we need only consider the behaviour of an effective 2LS to capture such effects in $\mathcal{H}_2$, which we write
$H_{2\rm{eff}} = \Delta\kb{\psi_+}{\psi_+}$ (in the basis $\{\ket{11},\ket{\psi_+}\}$). The system is coupled strongly to a laser mode and weakly to a reservoir of empty modes (the radiation field). In contrast to Section~\ref{model}, we treat the laser mode quantum mechanically, assuming it to be monochromatic and in a coherent state with mean photon number  $\exx{N} \gg \Delta N \gg 1$. We also demand that the number of photons emitted into new modes as a result of fluorescence is much smaller than the width of the photon distribution $\Delta N \gg \Gamma T$, with $T$ being a characteristic dot-light interaction time. Under these assumptions, we can  write the coupling term simply as $\Omega (\kb{11}{\psi_+} + \mathrm{h.c.}) / \sqrt{2}$~\cite{cohen-tannoudji92}. This form agrees with the coupling term for a classical laser (within the RWA), and so we formally retrieve the relevant terms in Eq. (\ref{eqn:h2sub_ham_rwa}).

\begin{figure*}
\begin{center}
\includegraphics[width=0.65\textwidth]{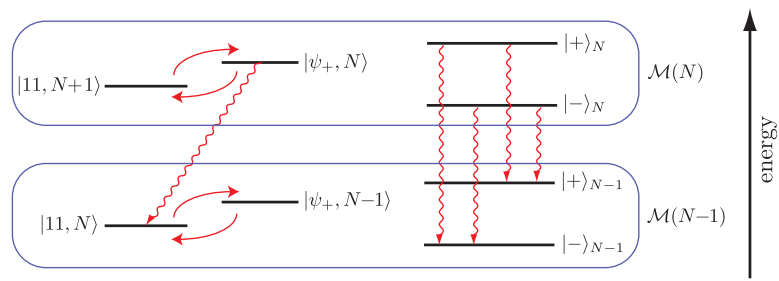}
\caption{Radiative decay from manifold $\mathcal{M}(N)$ to $\mathcal{M}(N-1) $. \textbf{Left:} the uncoupled basis. The energy difference between the two states in each manifold is the detuning $\Delta$, the solid arrows correspond to absorption and stimulated emission processes, and the wavy arrows denote spontaneous emission. \textbf{Right:} Allowed spontaneous emission transitions between the dressed states. The energetic splitting in each manifold is the effective Rabi frequency $\Omega' = \sqrt{\Delta^2 + 2\Omega^2}$ and the spacing between adjacent manifolds is the laser frequency $\omega_l$. The leftmost emission process is at frequency $\omega_l + \Omega'$, the two centre-lines emit at the frequency of the laser and the rightmost emission is at $\omega_l - \Omega'$; this gives rise to the famous Mollow triplet~\cite{mollow69}.}
\label{fig:dressed_state_transitions}
\end{center}
\end{figure*}

In absence of either the dot-laser coupling term or an environment, the energies of the combined QD and laser mode states $\{\ket{11, N+1}, \ket{\psi_+, N}\}$ differ by the detuning $\Delta \equiv \omega_0 - \omega_l+V_F$. If the detuning is small ($\Delta \ll \omega_0$), the states are energetically close to each other and can be conveniently  grouped into a manifold $\mathcal{M}(N)$. Likewise, we group the pairs $\{\ket{11, N}, \ket{\psi_+, N-1}\}$, $\{\ket{11, N+2}, \ket{\psi_+, N+1}\}$ etc. into a ladder of manifolds differing in energy by steps $\omega_l$. Introducing the dot-laser coupling mixes the bare eigenstates, again defining a dressed basis
\begin{eqnarray}
\ket{-}_N&{}={}&\cos \Theta \ket{11, N+1} - \sin\Theta \ket{\psi_+, N}, \label{eqn:dressed_minus} \\
\ket{+}_N&{}={}&\sin \Theta \ket{11, N+1} + \cos \Theta \ket{\psi_+, N}, \label{eqn:dressed_plus}
\end{eqnarray}
with $ \Theta=(1/2) \arctan{(\sqrt{2}\Omega/ \Delta)}$ as before, and we use the subscript $N$ to denote the manifold $\mathcal{M}(N)$. Recall the similar set of eigenstates $\ket{-} = \cos \Theta \ket{11} - \sin\Theta \ket{\psi_+} $, $\ket{+} = \sin \Theta \ket{11} + \cos \Theta \ket{\psi_+}$ that would be obtained by considering a classical laser. In contrast to the dressed states of Eqs. (\ref{eqn:dressed_minus}) and (\ref{eqn:dressed_plus}), the classical treatment does not give any information on the number of photons in the laser mode nor on the manifold $\mathcal{M}(N)$. As a result, it appears at first glance that \ket{-} is truly the system's ground-state from which no relaxation is possible. Yet, it is clear that for $\Theta \neq 0$ both dressed states have some excitonic character and a transition from each into either state of the adjacent manifold below, $\mathcal{M}(N-1)$, is possible. During this process a photon from the laser mode is emitted into a different, previously empty mode, by means of spontaneous emission (see Fig.~\ref{fig:dressed_state_transitions}). As is easily seen in the dressed picture, three transitions are possible, emitting photons at frequencies $\omega_l$, and $\omega_l \pm \Omega'$, where the effective Rabi frequency $\Omega' = \sqrt{\Delta^2 + 2\Omega^2}$ is the energy separation between the dressed states in each manifold. In the context of the CPHASE gate, this implies that following the state $\ket{\zeta_-}\approx\ket{-}$ adiabatically does not leave the operation immune to radiative decay. On the contrary, if the pulse is tuned all the way to resonance, $\Theta = \pi / 4$, half the population will be susceptible to recombination, analogous in fact to the case of resonant excitation where half the population remains in state \ket{11} and half in state \ket{\psi_+} on average.

To estimate the impact of spontaneous emission on the gate fidelity we need to know the transition rates between the dressed states. Since our proposal relies on adiabatic following of the $\ket{-}$ state, only its decay rate $\Gamma_-$ needs to be considered. It is given by the sum of two processes: $\ket{-}_N \to \ket{-}_{N-1}$ and $\ket{-}_N \to \ket{+}_{N-1}$. The total decay rate is then
\begin{equation}
\Gamma_-=\sqrt{2}\Gamma_0\sin^2 \Theta \cos^2 \Theta + \sqrt{2}\Gamma_0\sin^4 \Theta= \sqrt{2}\Gamma_0 \sin^2 \Theta,
\label{eqn:minus_decay}
\end{equation}
confirming $\Gamma_-=\Gamma_0 / \sqrt{2}$ for $\Theta = \pi / 4$, as expected, where $\Gamma_0$ is defined as the single dot spontaneous emission rate. We note that $\Gamma_-$ decreases as $\Theta$ becomes smaller, suggesting that keeping $\Theta$ small during the gating operation should be advantageous. However, as it is the energetic difference between the unperturbed \ket{00} state and the $\ket{-}$ states in $\mathcal{H}_1$, $\mathcal{H}_{1'}$, and $\mathcal{H}_2$ that leads to the CPHASE operation, a smaller mixing angle entails prolonging the gating time.
The interesting case to consider is that of weak mixing, $\Theta \ll \pi/2$, realized by keeping  $\Omega/ \Delta$ small. Rewriting Eq. (\ref{eqn:minus_decay}) as
\begin{equation}
\Gamma_- = \frac{\Gamma_0}{\sqrt{2}} \left( 1 - \frac{1}{\sqrt{1 + (\sqrt{2}\Omega / \Delta)^2}} \right)
\label{eqn:minus_decay_2}
\end{equation}
and expanding in a Taylor series around $\Omega/ \Delta = 0$, yields
\begin{equation}
\Gamma_- =  \frac{\Gamma_0}{\sqrt{2}} \left( \frac{\Omega}{\Delta} \right)^2 + \mathcal{O} \left(\frac{\Omega}{\Delta}  \right)^4.
\label{eqn:minus_decay_3}
\end{equation}
For fixed $\Omega$ we see that $\Gamma_- \sim \Delta^{-2}$ to leading order.

\begin{figure}
\begin{center}
\includegraphics[width=0.55\textwidth]{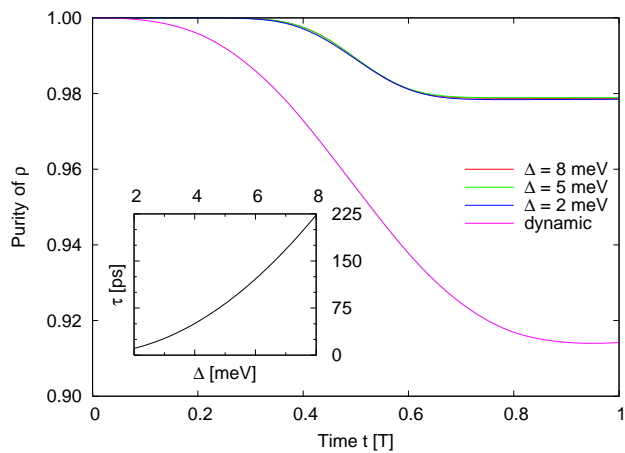}
\caption{Purity of the density matrix for a CPHASE gate with $\Gamma_0 = 1/100~\rm{ps}^{-1}$, $V_{XX} = 5$~meV and  $V_F = +0.85$~meV.  The dynamic gate uses  $\Omega = 0.1$~meV, whereas the adiabatic gates are performed at $\Omega_0 = 1$~meV with differing $\Delta$ (given in meV). The purity of the adiabatic operation does not depend on the detuning despite the substantial differences in the pulse duration, which are  shown in the inset. This is due to the clearly discernible quadratic dependence of $\tau$, the Gaussian pulse duration, on $\Delta$ (see text).}
\label{fig:2qd_optical_me}
\end{center}
\end{figure}

The CPHASE gate requires that the phase accumulated on \ket{11} exceeds that on \ket{01} and \ket{10} by $\pi$, the build-up being due essentially to $V_F$. During the pulse, \ket{11} follows \ket{\zeta_-} with an energy well approximated by $E_{\zeta_-} = (\Delta - \sqrt{\Delta^2 + 2 \Omega^2})/2$, whereas \ket{01} and \ket{10} follow the dressed eigenstates in their respective subspaces, each of which has an energy $E_{D} = (\Delta - V_F - \sqrt{(\Delta - V_F)^2 + \Omega^2})/2$. Again, in the limit of a large detuning, we Taylor expand  $E_{\zeta_-}$ around $\Omega / \Delta = 0$, yielding
\begin{equation}
E_{\zeta_-} = \frac{\Omega}{2} \left( \frac{\Omega}{\Delta} \right) +  \mathcal{O} \left( \frac{\Omega}{\Delta} \right)^3.
\end{equation}
Similarly, expanding $E_{D}$ around $\Omega / (\Delta -V_F) = 0$ gives
\begin{equation}
E_{D} = \frac{\Omega}{4} \left( \frac{\Omega}{\Delta -V_F} \right) +  \mathcal{O} \left( \frac{\Omega}{\Delta - V_F} \right)^3.
\end{equation}
Therefore, assuming $\Omega$ is fixed for the moment and that phase accumulates as $\phi=\delta E \,t$, we obtain to leading order 
\begin{equation}
\delta E = E_{\zeta_-} - 2 E_{D} = -  \frac{\Omega^2 V_F}{2 \Delta( \Delta - V_F)},
\end{equation}
for the relevant energy shift during the CPHASE operation.
For $\Delta \gg V_F$ the denominator is clearly dominated by $\Delta^2$, such that the required gating time $t$ would simply take the form $t = \pi / (E_{\zeta_-} - 2 E_{D}) \sim \Delta^2 / V_F$, in the case of a square pulse. Of course, the adiabatic gate relies on a Gaussian pulse profile, though in the limit $\Delta\gg\Omega$ we would expect the above analysis to provide a good approximation. Further, the numerical data in the inset of Fig. \ref{fig:2qd_optical_me} illustrate that the Gaussian pulse duration $\tau$ does also follow a parabolic form with $\Delta$. Comparing with $\Gamma_- \sim \Delta^{-2}$ from before, we see that any decrease in the decay rate $\Gamma_-$ by means of a larger detuning $\Delta$ is compensated for by the longer duration of the operation. 
Hence, contrary to the naive expectation, in this parameter regime applying a farther-detuned laser pulse does not improve the robustness of the adiabatic CPHASE gate to spontaneous emission.

To validate this result we perform a numerical simulation of the laser-driven CPHASE operation  [as described by Eqs. (\ref{eqn:pulse_detuning}) and (\ref{eqn:pulse_coupling})] accounting for radiative decay. 
The appropriate ME describing spontaneous emission from a 2LS dressed by a laser mode is \cite{cohen-tannoudji92}
\begin{equation}
\dot{\chi} = - i [H, \chi] + \Gamma\left(\sm \chi \sp - \frac{1}{2} (\sp \sm \chi + \chi \sp \sm)  \right),
\end{equation}
where $H = H_S + H_L + V$ is the joint Hamiltonian of the QD ($H_S$), the laser mode ($H_L$), and the interaction between them ($V$). Here, $\chi$ is the joint QD-laser density matrix and $\sp$ and $\sm$ are, respectively, the raising and lowering operators of the 2LS. Any thermally induced emission and absorption processes are neglected as the bosonic thermal occupancy $N(\omega_0)$ is extremely small even at room temperature. Information about the state of the laser mode is no longer needed, allowing us to sum over all manifolds and obtain a ME for the density matrix $\rho$ of the system alone. For the coherent part of the dynamics, we again retrieve the same form of $H_{2\rm{LS}}$ as the qubit Hamiltonian $H_S$ \cite{cohen-tannoudji92}. The incoherent, spontaneous emission part is described by the operator $\sm$ with associated rate $\Gamma$ \cite{breuer02,kowalewska01}. For $\rho$ the ME can be written in a compact way as \cite{hohenester04a}
\begin{equation}
\dot{\rho} = - i [H_{\rm nh}, \rho] + L_{sp} \rho L_{sp}^{\dagger},
\label{eqn:qoptical_me}
\end{equation}
where $H_{\rm nh} = H_S - (i/2) L_{sp}^{\dagger}  L_{sp}$ is an effective non-Hermitian Hamiltonian, and $L_{sp} = \sqrt{\Gamma} \sm$ is a Lindblad operator for the spontaneous emission processes. Radiative recombination affects each dot separately in $\mathcal{H}_1$ and $\mathcal{H}_{1'}$, and the Lindblad operator describing spontaneous emission in both subspaces is therefore the same as it would be in the single dot case, $L = \sqrt{\Gamma_0} \kb{01}{0X}$ and  $L = \sqrt{\Gamma_0} \kb{10}{X0}$, respectively. In the larger subspace $\mathcal{H}_2$ transformation to the diagonal basis still gives a single effective Lindblad operator, $L = (\sqrt{2} \Gamma_0)^{1/2} \kb{11}{\psi_+}$, acting with respect to the Hamiltonian of Eq. (\ref{eqn:h2sub_ham_rwa})~\footnote{No differentiation between positive and negative values of $V_F$ is necessary throughout this section as the radiative decoherence channel is insensitive to the sign.}.

We now use Eq. (\ref{eqn:qoptical_me}) to simulate the adiabatic CPHASE gate at the same time comparing it with the dynamic operation proposed in Ref~\cite{nazir04}. We present our results as the purity of the full 9LS density matrix, including all decoupled subsystems, taking a general input state of the form $\ket{\phi} = \left( \ket{00} + \ket{01} + \ket{10} + \ket{11} \right) / 2$. The purity is defined as the trace of the squared density matrix $\mathrm{tr}[ \rho^2]$. By definition, it is equal to one as long as the system is in a pure state, and decays to $1/d$ for a totally mixed state, where $d$ is the dimension of the Hilbert space \cite{nielsen00}. Decoherence effects render the system into a mixed state, thereby indicating corruption of the output through a reduction of the purity.
Figure \ref{fig:2qd_optical_me} shows the results obtained for a natural excitonic lifetime of $0.1$ns. The adiabatic curves are plotted for $\Omega_0 = 1$~meV and for various $\Delta$. We note that they nearly coincide, clearly showing that the purity has no apparent dependence on the value of $\Delta$, as expected from our earlier simple analysis. We see also that the adiabatic gate retains a purity much closer to unity than the dynamic gate, the speed of which is limited by the conditions set out in Ref.~\cite{nazir04}.

\section{Interaction with phonons}

In contrast to an isolated atom, a QD is embedded in the macroscopic solid state matrix of its substrate. The equilibrium positions of the lattice ions are determined by the overall electrostatic potential of all charge carriers. As a consequence, the electronic state of the QD couples to the lattice vibrations (phonons) \cite{mahan00}. Three processes result from the interaction of an exciton with phonons in semiconductor QDs: pure dephasing, phonon emission, and phonon absorption.

For many systems, pure dephasing is dominant at very low temperatures \cite{pazy02,takagahara99,krummheuer02}. On excitation, an abrupt change in the charge configuration shifts the equilibrium lattice positions of surrounding ions and a phonon packet is emitted as a lattice relaxation occurs. This corresponds to a transfer of information about the electronic state of the QD into the environment and can be interpreted as a kind of environmental `which way' measurement of the charge qubit \cite{roszak06}, resulting in a loss of coherence and forcing the system into a mixed state. Pure dephasing is energy conserving and does not lead to a relaxation in the QD. It is also an intrinsically non-Markovian process \cite{roszak05} and its influence has been studied extensively for Rabi oscillations \cite{forstner03,machnikowski04}, absorption line shapes \cite{krummheuer02, takagahara99} and in the context of the spin-boson model \cite{pazy02}. Several authors have shown how pure dephasing may be eliminated if the coherent system excitation is slow enough such that all lattice ions adiabatically follow their equilibrium positions \cite{alicki04, roszak05, calarco03, machnikowski04, machnikowski06}.

On the other hand, phonon absorption and emission correspond to real transitions between energy levels in the QD. Absorption is only possible at finite temperatures but can become a substantial problem in semiconductor structures even for temperatures well below 10K. Much like spontaneous photon emission, phonon-mediated relaxation processes are possible at any temperature. 
To study these effects, we derive a Markovian ME from first principles. Since we are interested primarily in the limit of low frequency dynamics we neglect optical phonons, which are separated by a large energy gap of $30$~meV or more \cite{krummheuer02}. In this situation, deformation and piezoelectric coupling to longitudinal acoustic (LA) phonons are the dominant phonon decoherence mechanisms. The interaction between charge carriers and acoustic phonons is generically given by a sum over the phonon modes $\vec{q}$ \cite{mahan00}
\begin{equation}
H_{ep} = \sum\limits_{\vec{q}} M_\vec{q} \hat{\varrho}(\vec{q}) (a_\vec{q}  + \cre{a}_\vec{-q}).
\end{equation}
Here, $M_\vec{q}$ is the coupling element,
\begin{equation}
M_\vec{q}  =  \sqrt{\frac{\hbar}{2 \mu V  \omega_\vec{q}}} C_\vec{q},
\label{eqn:phcoupling}
\end{equation}
where $\mu$ is the mass density, $V$ the lattice volume and  $\omega_\vec{q}$ the phonon frequency, with wavevector $\vec{q}$. The coupling constant is denoted $C_\vec{q} = D \vec{q}$ for the deformation potential and  $C_\vec{q} = P$ for piezoelectric coupling, while $\hat{\varrho}$ is the charge density operator
\begin{equation}
\hat{\varrho}(\vec{q})  =   \sum\limits_{i, j} \cre{d}_j \an{d}_i \int \mathrm{d}^3 r e^{- i \vec{q} \cdot \vec{r}} \psi^{\dagger}_j(\vec{r}) \psi_i(\vec{r}),
\label{eqn:chargeoperator}
\end{equation}
with $\cre{d}_i, \an{d}_i$ the creation and annihilation operators of charge carrier $i$, and $\psi_i(\vec{r})$ the corresponding wavefunction. For strongly bound excitons with no wavefunction overlap between carriers on adjacent QDs, the coupling elements are diagonal and given by the difference of those for electron and hole, $M^e_\vec{q}$ and $M^h_\vec{q}$, respectively, multiplied by the Fourier transform of the electron and hole density operator $\mathcal{P}[\psi^{e}(\vec{r})]$ and $\mathcal{P}[\psi^{h}(\vec{r})]$:
\begin{equation}
M_\vec{q} \hat{\varrho}(\vec{q}) = \sum\limits_{i} (M^e_\vec{q} \mathcal{P}[\psi_i^{e}(\vec{r})] - M^h_\vec{q} \mathcal{P}[\psi_i^{h}(\vec{r})]) \cre{c}_i \an{c}_i.
\label{eqn:phonon_coupling_element}
\end{equation}
Here, $\cre{c}_i$ ($\an{c}_i$) denote excitonic creation (annihilation) operators. 

Consider the two-dot system Hamiltonian of Eq. (\ref{eqn:h2sub_ham_rwa}), once more neglecting the biexcitonic level \ket{XX}~\footnote{In fact, on resonance, this assumption requires $\abs{V_{XX} - 2 V_F} \gg \abs{\Omega}$. However, for a larger detuning, $\Omega$ can be larger as well, and we have checked numerically that excitations to the \ket{XX} level remain well below $10^{-5}$ for all parameters used in the following.}
\begin{equation}
H_2^{''}=(\Delta - 2 V_F) \kb{\psi_-}{\psi_-} + \Delta  \kb{\psi_+}{\psi_+}+\frac{\Omega}{\sqrt{2}}  (\kb{11}{\psi_+}  + \mathrm{h.c.}).
\label{eqn:h2sub_ham_rwa_3ls}
\end{equation}
The full Hamiltonian including a common bath of phonons is given by $H = H_2'' + H_B + H_I$, where the bath and interaction terms are given by
\begin{equation}
H_B = \sum \limits_\vec{q} \omega_\vec{q}  \cre{a}_\vec{q}  \an{a}_\vec{q},
\label{eqn:ph_ham_bath}
\end{equation}
and
\begin{equation}
H_I = \sum\limits_\vec{q} \left( g^{0X}_\vec{q} c_{0X}^\dagger c_{0X} + g^{X0}_\vec{q} c_{X0}^\dagger c_{X0}  \right) (\an{a}_\vec{q} + \cre{a}_\vec{-q}),
\label{eqn:2dot_phonon_int}
\end{equation}
respectively.
In the absence of \ket{XX} the phonon interaction term involves only single exciton levels, with $g^{i}_\vec{q} =  M^e_\vec{q} p^e_{\vec{q}}(\psi_i) - M^h_\vec{q} p^h_{\vec{q}}(\psi_i)$. If the wavefunctions $\psi_{0X}(\vec{r})$ and $\psi_{X0}(\vec{r})$ are of identical form, albeit centred at different positions $\pm \vec{d}$, we obtain, by the shift property of the Fourier transform,
\begin{eqnarray}
g^{0X}_\vec{q}  &=& e^{+i \vec{q} \cdot \vec{d}}  \left( M^e_\vec{q} \mathcal{P}[\psi^e(\vec{r})] - M^h_\vec{q} \mathcal{P}[\psi^h(\vec{r})] \right), \\
g^{X0}_\vec{q} &=& e^{-i \vec{q} \cdot \vec{d}}  \left( M^e_\vec{q} \mathcal{P}[\psi^e(\vec{r})] - M^h_\vec{q} \mathcal{P}[\psi^h(\vec{r})] \right).
\end{eqnarray}
We now diagonalize Eq. (\ref{eqn:h2sub_ham_rwa_3ls}) and write $H_I$ in the resulting basis $\{ \ket{\zeta_-}, \ket{\zeta_+}, \ket{\psi_-} \}$, with respective eigenenergies $(\Delta -  \sqrt{2 \Omega^2 + \Delta^2})/2, (\Delta + \sqrt{2 \Omega^2 + \Delta^2})/2$ and $\Delta - 2 V_F$. Moving to the interaction picture with respect to $H_2''+H_B$ (see Appendix), the system operators are ordered by their frequencies, giving
\begin{eqnarray}
\tilde{H}_I(t) &{}={}&  \frac{1}{2}  \sum\limits_{\omega' \in \{ 0, \Lambda, \Upsilon, \Xi \}} \left( P_{\omega'} e^{-i \omega' t}  +  P_{\omega'}^{\dagger} e^{i \omega' t}  \right)\nonumber\\
  &&\:{\times}\sum \limits_\vec{q}  (g^{0X}_\vec{q} \pm g^{X0}_\vec{q}) (\an{a}_\vec{q} e^{-i \omega_\vec{q} t} + \cre{a}_\vec{-q} e^{i \omega_\vec{q} t}),
  \label{eqn:h2_interactionpic_operators}
\end{eqnarray}
where $\pm \equiv + $ for $\omega' \in \{ 0, \Lambda \}$ and  $ - $ for $\omega' \in \{\Upsilon, \Xi \}$. Here,
\begin{eqnarray}
P_0 &=& \frac{1}{2}( \cos^2 \Theta \kb{\zeta_+}{\zeta_+} + \sin^2 \Theta \kb{\zeta_-}{\zeta_-} + \kb{\psi_-}{\psi_-})\\
P_\Lambda &=& - \frac{1}{2} \sin 2 \Theta \kb{\zeta_-}{\zeta_+}, \label{eqn:p_lambda} \\
P_{\Upsilon} &=& \cos \Theta \kb{\psi_-}{\zeta_+}, \label{eqn:p_upsilon} \\
P_{\Xi} &=& - \sin \Theta \kb{\zeta_-}{\psi_-}, \label{eqn:p_xi}
\end{eqnarray}
with frequencies
\begin{eqnarray}
\Lambda &=& \sqrt{2 \Omega^2 + \Delta^2}, \\
\Upsilon &=& 2 V_F - \frac{1}{2} ( \Delta - (\sqrt{2 \Omega^2 + \Delta^2} ), \label{eqn:upsilon_freq} \\
\Xi &=& \frac{1}{2}( \Delta + (\sqrt{2 \Omega^2 + \Delta^2} ) - 2 V_F. \label{eqn:xi_freq}
\end{eqnarray}
In this form, the phonon operators Eqns. (\ref{eqn:p_upsilon}, \ref{eqn:p_xi}) are only appropriate  for the detuned adiabatic gate with $\Delta > 2 \abs{V_F}$. Otherwise, care must be taken to ensure all frequencies are greater than zero. Around resonance and for $V_F > 0$, $\Xi$ switches sign (see Fig. \ref{fig:dressed_states_energy_h2}) and we need to use $P_{\Xi}^{\dagger}$ with associated frequency $\abs{\Xi}$ instead. On the other hand, for  $V_F < \Omega / \sqrt{2}$,  $\Upsilon$ switches sign requiring the redefinition of $P_{\Upsilon}$  to $P_{\Upsilon}^\dagger$ with associated frequency $\abs{\Upsilon}$.

A master equation for the system evolution is now derived in the usual way \cite{breuer02, gardiner00} by integrating the von-Neumann equation for the joint density matrix $R$ of the system and bath and tracing over the phonon modes. This results in a reduced system density matrix obeying
\begin{equation}
\dot{\rho} = - \int\limits_0^t \mathrm{d}t' \mathrm{tr}_{ph} \left( [\tilde{H}_I(t), [\tilde{H}_I(t'), R(t')]] \right).
\end{equation}
The Born-Markov approximation is now performed, which relies on two assumptions. First, we assume there is no back-action from the small system on the much larger bath, meaning the joint density matrix can be written as a product at all times $R = \rho \otimes \rho_B$. Second, the bath has no memory and we therefore replace $\rho(t')$ by $\rho(t)$, justified for rapid bath relaxation \cite{breuer02,gardiner00}. If the system dynamics occurs on a timescale much faster than relaxation due to interactions with the bath, we can perform a RWA~\footnote{
The RWA requires $\Lambda, \omega'  \gg  J_{\pm}(\Lambda), J_{\pm}(\omega')$ the diagonal Lindblad-type ME  only being strictly valid in this limit. Fortunately, the RWA assumption is indeed justified when $\Omega_0 < 1$ and $\Delta = 0$ as in the dynamic case or $\Delta \gtrsim \omega_{e/h}$ as will be used in the adiabatic case.} to arrive at an interaction picture ME in Lindblad form \cite{breuer02, stace05}
\begin{eqnarray}
\dot{\rho}&{} ={}& J_+(\Lambda)\left( [N(\Lambda) + 1] D[P_{\Lambda}]\rho +  N(\Lambda) D[P_{\Lambda}^{\dagger}]\rho \right)\nonumber \\
&&\:{+}\sum\limits_{\omega' \in \{ \Upsilon, \Xi \}} J_-(\omega')\left( [N(\omega') + 1] D[P_{\omega'}]\rho +  N(\omega') D[P_{\omega'}^{\dagger}]\rho \right),
\label{eqn:ph_me_2dots}
\end{eqnarray}
with $D[L]\rho \equiv  L \rho L^{\dagger} - 1/2 (L^{\dagger} L \rho + \rho L^{\dagger} L)$ usually referred to as the `dissipator' of the ME. Here, \mbox{$N(\omega) = \left(\exp (\omega / k_B T) - 1 \right)^{-1}$} describes the thermal occupation of the phonon modes and $J_{\pm}(\omega)$ are the spectral densities given by 
\begin{eqnarray}
J_+(\omega)&{} ={}& 2 \pi \sum \limits_{\vec{q}} \abs{g_\vec{q}^{0X}+g_\vec{q}^{X0}}^2 \delta (\omega - \omega_\vec{q})
\label{eqn:spectralplus}\\
J_-(\omega)&{} ={}& 2 \pi \sum \limits_{\vec{q}} \abs{g_\vec{q}^{0X}-g_\vec{q}^{X0}}^2 \delta (\omega - \omega_\vec{q}).
\label{eqn:spectralminus}
\end{eqnarray}
Note that the Lindblad operator $P_0$ has been dropped since the spectral density vanishes in the limit of $\omega = 0$. Consequently, such a ME does not describe non-Markovian pure dephasing \cite{pazy02, krummheuer02} effects in the dressed basis, on account of the Born-Markov approximation, but does provide a suitable description of phonon-assisted transitions. In the bare basis, the Lindblad operators  $P_{\omega'}$ and $P_{\Lambda}$ consist of emission, and absorption terms and $P_{\Lambda}$ features an additional dephasing term.

\begin{figure}
\begin{center}
\includegraphics[width=0.55\textwidth]{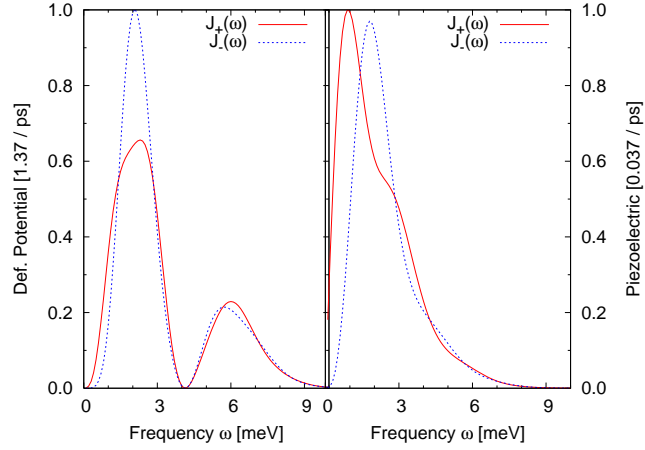}
\caption{Normalised spectral densities $J_{\pm}(\omega)$: left, for deformation potential, and right, for piezoelectric coupling (note the different scales for the $y$ axes of the two plots).}
\label{fig:spectral_densities_2dots}
\end{center}
\end{figure}

\begin{table}%[htdp]
\begin{center}\begin{tabular}{lr}
Electron deformation potential $D_e$	& 14.6 eV \\
Hole deformation potential $D_h$		&  4.8 eV \\
Piezoelectric coupling constant		& 1.45 eV / nm\\
Effective electron mass $m_e$			& 0.067 $m_0$ \\
Effective hole mass $m_h$			& 0.34 $m_0$\\
Mass density $\mu$					& 5.3 g / cm$^3	$\\
Velocity of sound $c_s$				&$ 4.8 \times 10^5$ cm / s
\end{tabular}
\caption{Material parameters for GaAs \cite{krummheuer02, pazy02, stace05}.}
\label{tab:GaAs_material_params}
\end{center}
\end{table}

For the calculation of $J_{\pm}(\omega)$ we choose the simple case of an isotropic harmonic confinement potential for each dot. The excitonic ground-state therefore has a Gaussian wavefunction \cite{nazir05}
\begin{equation}
\psi_{e/h}(\vec{r}) = \left( \frac{1}{d_{e/h} \sqrt{\pi}}   \right)^{\frac{3}{2}} e^{- \frac{r^2}{2 d^2_{e/h}}},
\label{eqn:gaussian_wavefunction}
\end{equation}
with $d_{e/h} = (\hbar / \sqrt{m_{e/h} c})^{1/2}$ giving the width of the wavefunction envelopes, which depend on the different effective masses $m_{e/h}$ for electrons and holes. The confinement strength $c$ is chosen to be $c = 8.3 \times 10^{-3} \mathrm{J/m}^2$ such that electrons and holes are subject to a harmonic potential  $V = c r^2$  with strength $162$~meV at $r = 2.5$~nm from the centre of the dot.
Assuming a linear and isotropic phonon dispersion $\omega_\vec{q} = c_s \abs{\vec{q}}$ we then obtain for the deformation potential
\begin{eqnarray}
J_{\pm}(\omega)&{} ={}& \frac{D_e^2 \hbar }{4 \pi \mu c_s^2 R^3} \left(\frac{\omega}{\omega_p} \right)^3 \left(1 \pm \mathrm{sinc} ~ \frac{ \omega}{  \omega_p} \right)\nonumber\\
&&\:{\times}\left( e^{-(\omega / \omega_{e})^2} - 2 \frac{D_h}{D_e} e^{-(\omega / \omega_{eh})^2}  + \frac{D_h^2}{D_e^2} e^{-(\omega / \omega_{h})^2} \right),
\end{eqnarray}
where $\omega_p = c_s /  R$. Both spectral densities are obviously of superohmic form~\cite{leggett87}, with high-frequency cutoff terms $\omega_{e,h} = \sqrt{2} c_s / d_{e,h}$ and $\omega_{eh} = 2 c_s / \sqrt{d_e^2 + d_h^2} $ related to the finite QD size. These terms filter out phonons with wavelengths too short to interact with the dots.
For piezoelectric coupling we get ohmic spectral densities given by
\begin{eqnarray}
J_{\pm}(\omega)&{} ={}& \frac{\hbar P^2}{4 \pi \mu c_s^2 R} \frac{\omega}{\omega_p}  \left(1 \pm \mathrm{sinc} ~ \frac{ \omega}{  \omega_p} \right) \nonumber \\
&&\:{\times}\left( e^{-(\omega / \omega_{e})^2} - 2 e^{-(\omega / \omega_{eh})^2}  + e^{-(\omega / \omega_{h})^2} \right).
\end{eqnarray}
All four spectral functions are plotted in Fig. \ref{fig:spectral_densities_2dots} for the parameters given in Table \ref{tab:GaAs_material_params}, and with a dot centre-to-centre distance of $R =  7$~nm, fulfilling the assumption of negligible wavefunction overlap. Crucially, the dots are still close enough to sustain a significant F\"orster interaction and dipolar shift~\cite{nazir05}. From Fig.~\ref{fig:spectral_densities_2dots} we identify deformation potential coupling as the dominant phonon-decoherence mechanism and neglect piezoelectric coupling in the following. This agrees with the literature \cite{pazy02, krummheuer02} and is primarily due to the fact that electrons and holes couple individually and with a different strength to phonons through the deformation potential, whereas only the difference in their wavefunctions contributes to the piezoelectric coupling. 
The difference in deformation potentials $D_e$ and $D_h$ combined with the slightly different electron and hole cutoff frequencies gives rise to the two peaks in the left hand side of Fig. \ref{fig:spectral_densities_2dots}. The drop down to zero between the peaks is a `sweet spot' at which point the phonon couplings completely cancel out.
Typically, self-assembled QDs are disc-shaped with a lateral width of $10 - 20$ nm and a smaller height of around $5$ nm in the growth direction. Accounting explicitly for this would introduce an angular dependency into the spectral density with the largest value of the cut-off frequency being in the growth direction, and having a value similar to that obtained using the isotropic potential model\footnote{Accounting for the anisotropy would also slightly affect the position of the minima of the deformation potential spectral density in Fig. \ref{fig:spectral_densities_2dots}, which would cease to be identically zero after averaging over the angular dependence.}. Hence, the error introduced by the simple model above is likely to be an overestimation of the magnitude of the spectral density towards large frequencies, with no considerable change in the position of the cutoff. Furthermore, in the limit of very small frequency, the spectral density does not depend on the form of the wavefunctions for electron and hole at all because the Fourier transforms then give $\mathcal{P}[\psi^{e/h}(\vec{r})] \approx 1$. 

\begin{figure}
\begin{center}
\includegraphics[width=0.55\textwidth]{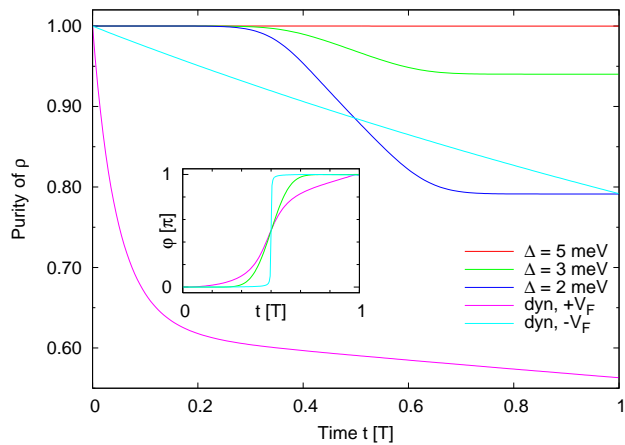}
\caption{The purity of the full density matrix during a CPHASE operation at $T = 5$~K. The system is coupled to a phonon bath as described by the ME (\ref{eqn:ph_me_2dots}). For the dynamic operation, we show results for $V_F = \pm 0.85$~meV, as indicated by the labels ``dyn, $+V_F$'' and ``dyn, $-V_F$''. The phase build-up of the dynamic gate (shown in the inset) should be step-like but is considerably smeared out in the case of the positive $V_F$, indicative of the fact that this approach  does not work (see text). The congruence of the terminal value of the adiabatic curve for $\Delta=2$~meV and the dynamic curve with negative $V_F$ is coincidental.}
\label{fig:two_dot_phonons}
\end{center}
\end{figure}

Solving Eq. (\ref{eqn:ph_me_2dots}) allows us to characterise the performance of both adiabatic and dynamic operation of the CPHASE gate in the presence of a phonon bath. We find that the dynamic gate, with positive coupling $V_F = 0.85$~meV, fares poorly even at zero temperature due to phonon emission processes. In this case, \ket{\psi_-} is the lowest energy eigenstate at resonance such that relaxation from both \ket{\zeta_-} and \ket{\zeta_+} into the dark state \ket{\psi_-} is possible. By the definitions of $\Upsilon$ and $\Xi$ [Eqs. (\ref{eqn:upsilon_freq}) and (\ref{eqn:xi_freq})], it is clear that a reduction of both these frequencies simultaneously to values much smaller than $2V_F$ is impossible. The result is that phonon emission processes, being roughly proportional to $J_{\pm} (2V_F)$, always operate at a fast rate, eventually transfering all population initially in \ket{11} into \ket{\psi_-}. However, if the F\"orster coupling is negative, with $ \abs{V_F}  > \Omega / \sqrt{2}$, the upwards shift of  \ket{\psi_-} is sufficiently large to lift it above both \ket{\zeta_-} and \ket{\zeta_+} at resonance and the dynamic scheme then works relatively well at low temperatures. In contrast, in the adiabatic approach phonon emission is always prevented for $\Delta > 2 \abs{V_F}$ because \ket{\zeta_-} is then the ground state of the system, yielding a perfect gating operation at zero temperature. In fact, if $V_F$ is negative this is even true irrespective of its particular value. 
At finite temperature we expect the adiabatic performance to improve with increasing detuning due to the rapid reduction in spectral density for large frequencies.

In Fig. \ref{fig:two_dot_phonons}, we plot the purity of the full 9LS density matrix at $T = 5$~K for both the dynamic and adiabatic approaches. The purity of the dynamic gate with positive F\"orster coupling drops quickly to a value of around $0.62$, corresponding to the expected state where all population has been transferred into \ket{\psi_-}. Of course, decoherence also occurs in $\mathcal{H}_1$ and $\mathcal{H}_{1'}$,  and this is included in the calculation as reflected by the slow and steady decline of the curve as time progresses. As expected, the performance of the adiabatic gate improves dramatically with increasing $\Delta$ by pushing both $\Lambda$ and $\Xi$ beyond the spectral cutoff \footnote{ $\Upsilon$ does not increase with larger $\Delta$, but this turns out to be unimportant as $P_{\Upsilon}$ couples states which, ideally, remain unpopulated.}, meaning $J(\Lambda)$ and $J(\Xi)$ decay exponentially. Physically, this implies that phonons with an energy matching the transitions between the relevant system eigenstates no longer interact with the dot because of their short wavelengths. 
This advantageous regime sets in at an energy below $10$~meV, allowing us safely to assume a linear phonon dispersion and to ignore optical phonon modes.

\section{Landau Zener transitions}
\label{section:LZ}

Non-adiabatic LZ transitions between system eigenstates are a further error source in the adiabatic scheme and can only be avoided by varying the control parameters more slowly.
There is, of course, a limit as to how slowly things can be done, since typically many operations need to be performed within the lifetime of the electron spin (roughly a millisecond~\cite{kroutvar04}).

\begin{figure}
\begin{center}
\includegraphics[width=0.55\textwidth]{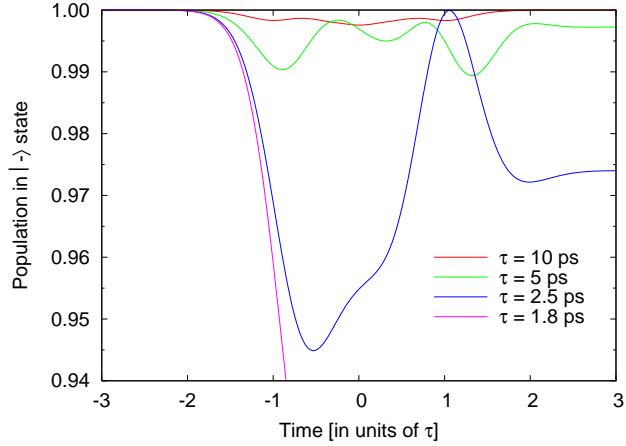}
\caption{The population remaining in the \ket{-}  ($\approx \ket{\zeta_{-}}$) state of the $\mathcal{H}_2$ subspace 
is shown for different pulse durations $\tau$ of a laser pulse $\Omega(t) = \Omega_0 \exp [-(t / \tau)]$ and $\Omega_0 = \Delta = 1~\rm{ps}^{-1}$. In this case the  adiabaticity condition (\ref{eqn:adiabaticity_condshort}) requires $\tau \gg \sqrt{2}~\rm{ps}$. The final population leakage for the $\tau = 1.8$~ps curve is roughly 8.5 percent and off the scale of this plot.}
\label{fig:lz_transition_amplitude}
\end{center}
\end{figure}

For the simplest case of a 2LS with an avoided crossing, the final transition amplitude was derived by Landau, Zener and St\"uckelberg \cite{landau32,zener32,stueckelberg32}. Unfortunately, this theory cannot generally be applied to driven qubits so that obtaining the transition amplitude often involves numerically solving the Schr\"odinger equation. Wubs \textit{et al.} \cite{wubs05} performed a detailed study of LZ transitions in optically controlled qubits, including subtleties such as phase effects of the driving laser.
They find that if the final transition amplitude vanishes after the pulse has finished, i.e. all population returns to the ground state, there may still be a significant population transfer at intermediate times. In particular, for  a symmetric laser pulse and a transition amplitude not much larger than one in a thousand at all times during the operation, almost the entire population undergoing a LZ transition is transferred back into the \ket{-} state once the pulse has finished. This is illustrated in some of the curves of Fig. \ref{fig:lz_transition_amplitude}. 
This is undesirable as it then becomes difficult to accurately predict the final phase achieved during the adiabatic operation. Furthermore, any population transferred to the excited state is much more susceptible to decoherence processes.

In general, any time dependent perturbation of a 2LS will lead to a finite transition probability between the dressed states. In order to keep this probability sufficiently small, we must explore the conditions under which adiabatic following can reasonably be expected. An adiabaticity condition for the simple model of a linear detuning sweep $\Delta = \dot{\Delta} t$ and constant coupling strength $\Omega / \dot{\Delta} \ll 1$ has been suggested in Refs.~\cite{calarco03,lovett05}. Here, we take a more general approach and assume that both $\Delta$ and $\Omega$ are time-dependent unless explicitly stated.

\begin{figure*}
\begin{center}
\includegraphics[width=0.495\textwidth]{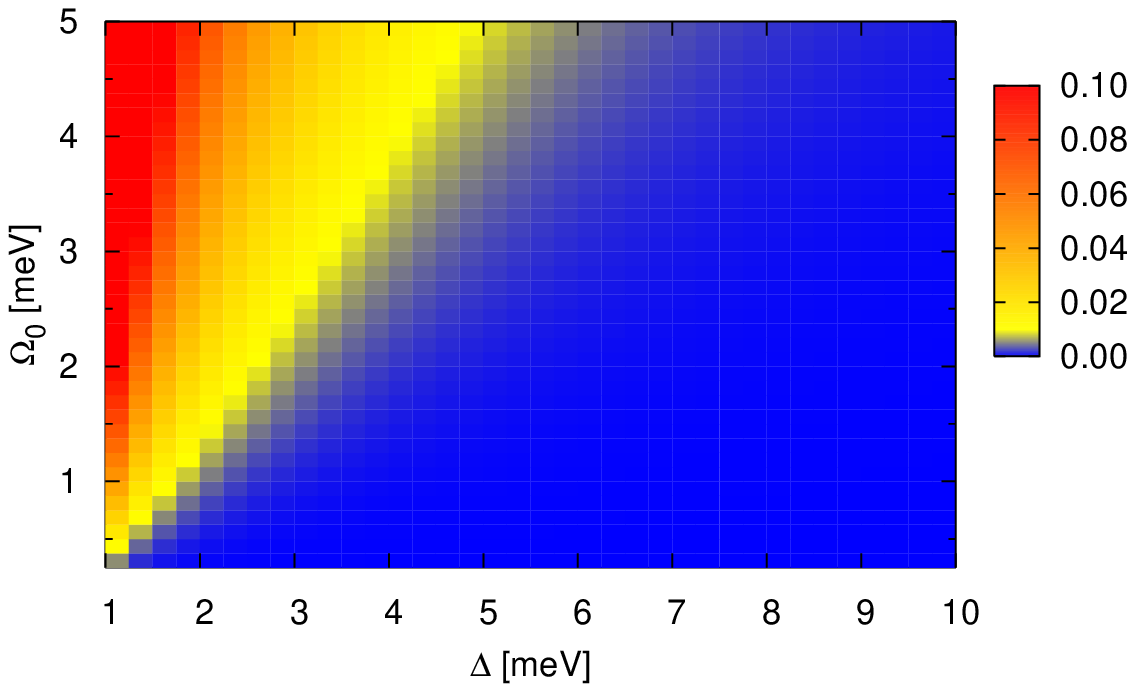}
\includegraphics[width=0.495\textwidth]{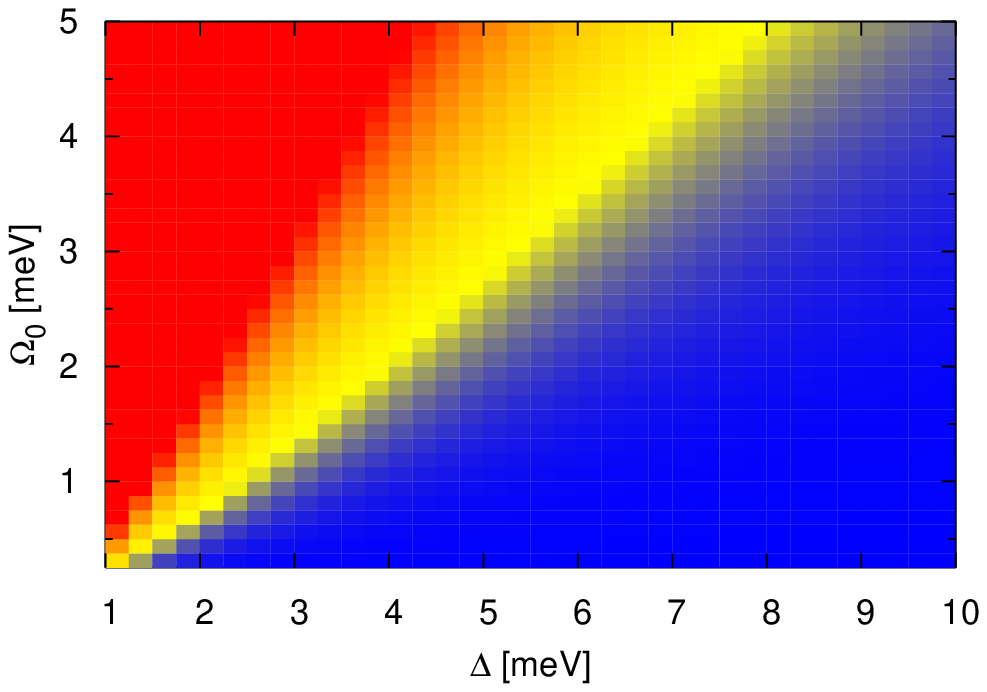}
\caption{Comparison of adiabaticity conditions. \textbf{Left:} the left hand side of inequality (\ref{eqn:adiabaticity_condition}) is plotted as a function of detuning and coupling strength. The pulse duration $\tau$ is chosen to generate a phase shift of $\pi$. The blue colour denotes a region where LZ transitions are adiabatically suppressed,  i.e. where the LHS of (\ref{eqn:adiabaticity_condition}) is much smaller than one. Towards the yellow region, it approaches values of $1/100$ and the approximation starts to break down. The red colour denotes a value of $1/10$ or more and adiabatic following can no longer be expected.  \textbf{Right:} the same for the LHS of Eq. (\ref{eqn:adiabaticity_condshort}) divided by $\tau$. We observe the same qualitative behaviour as in the left part of the plot, however, this simplified form  of the adiabatic condition is more stringent than the previous one, confirming that it guarantees adiabatic following, but is not necessarily required for it.}
\label{fig:adiabaticity_condition}
\end{center}
\end{figure*}

To derive an adiabaticity condition for the CPHASE gate we again consider the 2LS approximation to Eq.(\ref{eqn:h2sub_ham_rwa}), valid as before for large positive detuning. In fact, in this regime the level spacing between \ket{\zeta_-} and the upper dressed state \ket{\zeta_X} is significantly larger than that between \ket{\zeta_-} and \ket{\zeta_+}. Furthermore, the coupling between \ket{\zeta_-} and \ket{\zeta_X} originates from the coupling of \ket{\psi_+} to \ket{XX}, but, at large detuning, \ket{\zeta_-} only contains a small admixture of \ket{\psi_+}. Therefore, the probability of LZ transitions from \ket{\zeta_-} to \ket{\zeta_X} is doubly small and it is sufficient to consider only those between \ket{\zeta_-} and \ket{\zeta_+}. The transformation to the basis of approximate instantaneous eigenstates (see Eq. (\ref{eqn:zetaapproxdressed}),
\begin{equation}
U =
\left(\begin{array}{cc}
\cos \Theta & - \sin \Theta \\
\sin \Theta & \cos \Theta
\end{array}\right),
\end{equation}
is time-dependent because $\Omega$ and $\Delta$ vary with time (recall that  $2 \Theta = \arctan (\sqrt{2}\Omega / \Delta)$), while the Hamiltonian transforms as
\begin{equation}
\tilde{H} = U^{\dagger} H U + i \left(  \ddt  U^{\dagger} \right) U.
\end{equation}
Let $\lambda^{\pm} = 1/2 (\Delta \pm \sqrt{\Delta^2 + 2\Omega^2})$ denote the approximate instantaneous eigenenergies of the states \ket{\zeta_-} and \ket{\zeta_+}. In this basis
\begin{equation}
\tilde{H} = \lambda^{-} \kb{\zeta_-}{\zeta_-} + \lambda^{+} \kb{\zeta_+}{\zeta_+} + \dot{\Theta} \left(i\kb{\zeta_-}{\zeta_+} + \mathrm{h.c.}  \right),
\label{eqn:2ls_hamiltonian_strictly_diagonalised}
\end{equation}
where the off-diagonal terms in $\tilde{H}$ couple the eigenstates and lead to LZ transitions. To achieve adiabatic following, the magnitude of these terms must be much smaller than the energetic difference between the eigenstates, giving
\begin{equation}
\dot{\abs{\Theta}} \ll \abs{\lambda^{+} - \lambda^{-}}.
\label{eqn:adiabaticity_inequality}
\end{equation}
This is equivalent to the more familiar condition $\bra{\zeta_-} \ddt \ket{\zeta_+} \ll\abs{\lambda^{+} - \lambda^{-}}$ \cite{wubs05}. Inserting $\lambda^{\pm}$ we arrive at the following adiabaticity condition
\begin{equation}
\frac{\dot{\Omega}\Delta - \Omega \dot{\Delta}}{\sqrt{2}(\Delta^2 + 2\Omega^2)^{3/2}} \ll 1,
\label{eqn:adiabaticity_condition}
\end{equation}
valid for arbitrary field amplitude $\Omega$ and detuning $\Delta$. If the temporal evolutions of $\Omega$ and $\Delta$ are known, this condition can be brought into the form $F(\Delta_0, \Omega_0) \ll \tau$, where $F(\Delta_0, \Omega_0)$ is a time-independent relation of known parameters and $\tau$ gives the characteristic time of the applied pulse. For instance, in the particular case considered here of a Gaussian field amplitude $\Omega = \Omega_0 \exp[-(t/\tau)^2]$ and constant detuning $\Delta = \Delta_0$, an upper bound for adiabaticity is found to be
\begin{equation}
\sqrt{2}\Omega_0 / \Delta_0^2 \ll \tau.
\label{eqn:adiabaticity_condshort}
\end{equation}
In Fig.~\ref{fig:adiabaticity_condition} we evaluate the adiabaticity conditions of both Eqs. (\ref{eqn:adiabaticity_condition}) and (\ref{eqn:adiabaticity_condshort}), plotting as a function of $\Delta$ and $\Omega_0$. The same conditions should also hold in subspaces $\mathcal{H}_1$ and $\mathcal{H}_{1'}$, with the change $\sqrt{2}\Omega\rightarrow\Omega$. We see clearly that LZ transitions do not occur in the limit $\Delta \gg \Omega_0$, exactly the same limit as that for which phonon transitions are suppressed, allowing us to circumvent both error sources simultaneously.

\section{CPHASE gate fidelity}

\begin{figure}
\begin{center}
\includegraphics[width=0.55\textwidth]{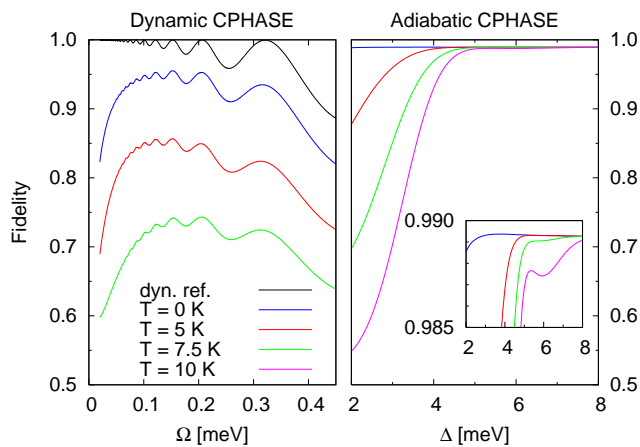}
\caption{Fidelity of the CPHASE gate. \textbf{Left:} fidelity of the dynamic operation for  a negative $V_F=-0.85$~meV. The dynamic reference curve gives the fidelity without additional decoherence (see text). \textbf{Right:} fidelity of the adiabatic operation for fixed $\Omega_0 = 1$~meV. The inset shows the same curves on a different scale to resolve more subtle features.}
\label{fig:2qd_fidelity}
\end{center}
\end{figure}

To bring together the results of the previous sections we now calculate the overall fidelity of the CPHASE operation, obtained from a numerical solution to a ME that includes both spontaneous emission as well as phonon-induced processes (and, of course, allows for LZ transitions). We take an input state given by $\ket{\phi_i} = (\ket{00} + \ket{01} + \ket{10} + \ket{11}) / 2$, which, after the CPHASE operation, should ideally produce an output state $\ket{\phi_f} = (\ket{00} + \ket{01} + \ket{10} - \ket{11}) / 2$. However, in practice, for both the adiabatic and dynamic approaches some phase is picked up in the $\mathcal{H}_1$ and $\mathcal{H}_{1'}$ subspaces. This needs to be ``unwound'' using two single qubit operations that we do not consider here. We therefore define the fidelity as $\mathcal{F} = \bra{\phi_f'} \rho \ket{\phi_f'}$, where \ket{\phi_f'} accommodates the additional phases on \ket{10} and \ket{01}, and $\rho$ is the full density matrix of the system after the gate has finished, including all detrimental environmental effects. As before, we use the material parameters of Table~\ref{tab:GaAs_material_params} and assume a single dot radiative decay rate of $\Gamma_0 = 0.01~\rm{ps}^{-1}$.

The results are plotted in Fig. \ref{fig:2qd_fidelity}. The left hand side shows the fidelity of the dynamic gate with $V_F = -0.85$~meV as a function of the coupling strength $\Omega$, and for different temperatures. The degradation of the fidelity due to detuned Rabi oscillation in both $\mathcal{H}_1$ and $\mathcal{H}_{1'}$ is included and, as expected, the effect is oscillatory and becomes more pronounced as $\Omega$ increases~\cite{nazir04} (to further illustrate this point, we have omitted all other sources of decoherence for the curve labelled ``dyn. ref.''). The fidelity is limited by the finite excitonic lifetime towards small values of $\Omega$, while to the right of the plot phonon-induced processes become more important. The best fidelity achieved here is roughly $0.95$ at absolute zero, and decreases even further at finite temperature.

The contrasting behaviour of the adiabatic scheme is shown in the rhs of Fig. \ref{fig:2qd_fidelity}. Here, the gate fidelity is plotted as a function of the detuning $\Delta$, for a fixed value of $\Omega_0 = 1$~meV. In agreement with the conclusions of the previous sections, the fidelity can be substantially increased simply by applying a suitably large detuning to the driving laser. In this limit, the fidelity is restricted by spontaneous emission, as the effects of LZ and phonon-induced transitions are confined to relatively small values of $\Delta$. Nonetheless, for $\Delta>5$~meV the fidelity is greater than $0.985$ for all temperatures shown and would improve even further for a smaller spontaneous emission rate (the rate used here translates into a rather pessimistic excitonic lifetime of $0.1$~ns). For interest, the inset shows an intermediate peak only visible at higher temperatures, which is related to the dip in the spectral density of the deformation potential (see Fig.~\ref{fig:spectral_densities_2dots}). 

\section{Conclusion}

To summarise, we have performed a realistic decoherence study of both the adiabatic and dynamic approaches to exciton-mediated spin manipulation in coupled QDs. We have shown that while dynamic gates suffer from rapid decoherence at finite temperatures (to an extent that may prohibit fault-tolerant QC), performing off-resonant adiabatic manipulations allows us to greatly suppress decoherence during the entangling operations. 
While a trade-off situation to minimise overall decoherence arises for the coupled QD dynamic scheme, this is no longer the case in the adiabatic scheme. Since no upper bound is imposed on the gate duration by radiative decay, a detuned and slow adiabatic operation is suitable to alleviate the adverse effects of both phonon-induced decoherence and LZ transitions. 
In this case, it is the finite spin coherence time that sets a bound on the possible gate duration. However, the adiabatic gates we consider operate on timescales of around $100$~ps, much shorter than typical spin coherence times~\cite{kroutvar04} and only about an order of magnitude greater than the ``fast" dynamic gates. We therefore conclude that adiabatic optical manipulation is a remarkably robust method for entangling spin qubits embodied in semiconductor nanostructures.

\ack
We thank Avinash Kolli and Jay Gambetta for stimulating and interesting discussions. This work was supported by the Marie Curie Early Stage Training network ÔQIPESTÕ (MEST-CT-2005-020505) within the European Community 6th Framework Programme. The contents of this paper reflect the authorÕs views and not the views of the European Commission.  The work is also supported by the National Research Foundation and Ministry of Education, Singapore. We also thank the QIPIRC (No. GR/S82176/01) for support. AN is supported by Griffith University, the State of Queensland, the Australian Research Council Special Research Centre for Quantum Computation, and the {\sc EPSRC}. BWL and SCB acknowledge support from the Royal Society.
\appendix

\section{Interaction picture transformation}

The system interaction picture operators of Eq. (\ref{eqn:h2_interactionpic_operators}) are strictly valid when the system's eigenfrequencies are constant in time, as is the case for the dynamic control scheme. This Appendix gives a justification for using the same interaction picture operators within the adiabatic approach, which has a time-varying system Hamiltonian due to $\Omega(t)$. We start with a qualitative argument followed by a more formal justification.

The transformation of Eq.~(\ref{eqn:2dot_phonon_int}) to the interaction picture as in Eq. (\ref{eqn:h2_interactionpic_operators}) would be strictly correct if the system frequencies $\Lambda$, $\Upsilon$, and $\Xi$ were constant. We require each of them to change sufficiently slowly in time to achieve adiabatic following. As long as $\dot{\Lambda} t \ll \Lambda$, etc. $\Lambda$, $\Upsilon$, and $\Xi$ can be assumed to be essentially constant at any given moment of time, such that a system operator $P_{\omega'}$ is well described by $P_{\omega'} e^{i \omega' t}$ as its interaction picture representation (up to a quasiconstant additional phase only varying on timescales much larger than $1 / \omega')$). Since the Markovian approximation in the derivation of the ME only depends on the current time, it seems only reasonable and consistent to use an instantaneous interaction picture operator.

More formally, we want to transform
\begin{equation}
H(t) = H_S(t) + H_B + H_I
\end{equation}
to the interaction picture with respect to $H_S(t)$ and $H_B$. $H_S(t)$, $H_B$ and $H_I$ are given by Eqs.
(\ref{eqn:h2sub_ham_rwa_3ls}), (\ref{eqn:ph_ham_bath}) and (\ref{eqn:2dot_phonon_int}), respectively.
The precise interaction picture transformation for a constant $H_B$ and a time-dependent $H_S(t)$ is given by the unitary transformation
\begin{equation}
U(t) = T \exp \left( - i \int \limits_{0}^{t} \mathrm{d} \tau H_S(\tau) \right) e^{ - i H_B t},
\label{eqn:intpic_transformation_time_ordering}
\end{equation}
where $T$ is the time ordering operator and we have chosen the Schr\"odinger and the interaction picture to coincide at the arbitrarily chosen initial time $t =0$. As  $H_S(t)$ does not commute with itself at different times, we go into the diagonal basis [in a similar way as for Eq. (\ref{eqn:2ls_hamiltonian_strictly_diagonalised})] and perform the adiabatic approximation, so that the system Hamiltonian is simply
\begin{equation}
H'_S(t) = \lambda^{-}(t) \kb{\zeta_{-}}{\zeta_{-}} + \lambda^{+}(t) \kb{\zeta_{+}}{\zeta_{+}} + \lambda^{\psi}(t) \kb{\psi_{-}}{\psi_{-}},
\label{eqn:adiabatic_diagonalised_form}
\end{equation}
where the $\lambda^{\pm, \psi}(t)$ are functions of time. Hamiltonian (\ref{eqn:adiabatic_diagonalised_form}) now commutes with itself at different times and we can drop the time ordering operator from Eq. (\ref{eqn:intpic_transformation_time_ordering}).  The system part  $U_S(t)$ of the transformation is then
\begin{equation}
U_S(t) = \exp \left( - i \int \limits_{0}^{t} \mathrm{d} \tau H'_S(\tau) \right).
\label{eqn:intpic_transformation}
\end{equation}
Instead of using Eq. (\ref{eqn:intpic_transformation}) to transform into the interaction picture, we apply the following transformation
\begin{equation}
U_S(t) = \exp \left( - i H'_S(t) t \right)
\end{equation}
to obtain transformed system operators $P_{\omega'} e^{i \omega' t}$ with nearly constant $\omega'$, which is the form we require for the derivation of the Markovian ME. However, since the total  Hamiltonian transforms as
\begin{equation}
\tilde{H} = U^{\dagger} H U + i \left(  \ddt  U^{\dagger} \right) U,
\end{equation}
this gives
\begin{equation}
\tilde{H} = H_I -  \left( \ddt H'_S(t) \right) t,
\end{equation}
rather than the desired $\tilde{H} = H_I$ as the transformed Hamiltonian. The additional contribution can be neglected if  $\dot{H}'_S t \ll H'_S$  (this condition is essentially equivalent to the condition on the eigenfrequencies, $\dot{\omega'} t \ll \omega'$, as stated above). The ME derived in this way then fails to capture the effect of the phonon bath on a small part of the system's dynamics. Therefore, the error is likely proportional to the overall effect caused by the phonon bath rather than being a constant of a given magnitude. Most importantly, the approximation made by neglecting the additional contribution to $\tilde{H}$ improves rapidly as $\Delta$ increases relative to $\Omega$. In the limit $\Delta \gg \Omega$, for which we predict a very high fidelity, the approximation is then very good indeed.\\

\end{document}